\documentclass[12pt]{article}

\usepackage{amsmath,amssymb}
\usepackage{amsfonts}
\usepackage{graphicx}
\usepackage{a4wide}

\usepackage{srcltx}
\usepackage{amsmath}
\usepackage{amssymb}
\usepackage{array}
\usepackage{tabularx}
\usepackage{epsfig}
\usepackage{psfrag}
\usepackage{enumerate}
\usepackage{url}
\usepackage{mathrsfs}
\usepackage{dsfont} % instead of bbm !!!

\usepackage[mathscr]{euscript}
\usepackage{multirow}
\def\bfm#1{\boldsymbol{#1}}
\def\E{\operatorname{\mathbf E}\nolimits}
\def\P{\operatorname{\mathbf P}\nolimits}

\def\nbinom#1#2{\mbox{\small$\dbinom{#1}{#2}$}}

\usepackage[T2A]{fontenc}

\begin{document}

\noindent {\it Problems of Information Transmission} \\
\noindent Vol. 58, No. 3, pp. 3-17, 2022.

%\vskip 0.4cm

\begin{center}
{\large\bf On the Reliability Function for a BSC with Noiseless
Feedback at Zero Rate} \footnote[1]{The reported study was
funded by RFBR according to the research project 19-01-00364.}
\end{center}

\begin{center} {\bf M. V. Burnashev} \end{center}

\begin{center}{\it Kharkevich Institute for Information Transmission Problems,\\
Russian Academy of Sciences, Moscow, Russia \\
email}: burn@iitp.ru
\end{center}

{\begin{quotation} \small {\bf Abstract}--
We consider the transmission of nonexponentially many messages \\
through a binary symmetric channel with noiseless feedback.
We obtain an upper bound for the best decoding error exponent. Combined with
the corresponding known lower bound, this
allows to find the reliability function for this channel at zero rate.
\end{quotation}}
{\begin{quotation} \small
{\it Key words:} {reliability function, noiseless feedback}.
\end{quotation}}

\section{Introduction and Main Results}
The binary symmetric channel $\rm BSC(p)$ with crossover probability $0<p<1/2$,
$q=1-p$, and noiseless feedback is considered. We consider the
case when the overall transmission time $n$ and $M_n=2^{Rn}$, $0<R<1$,
equiprobable messages $\{\theta_1,\theta_2,\ldots,\theta_{M_n}\}$ are given.
After the moment $n$ the receiver makes a decision ${\hat \theta}$ on the true
message $\theta_{\rm true}$ transmitted.

Define the minimal possible decoding error probability
\begin{equation}\label{deferprob}
P_{\rm e}(M_n,n,p)=\min\frac{1}{M_n}\sum\limits_{i=1}^{M_n}
P({\rm e}| \theta_i),
\end{equation}
where $P({\rm e}| \theta_i)$ -- conditional decoding error probability for a transmission method used, provided $\theta_i$ is the true message
$\theta_{\rm true}$, and minimum is taken over all transmission methods of
length $n$.

Denote by $F(R,p)$, $0<R<1$, the best decoding error exponent for
$M_n=e^{Rn}$ codewords over $\rm BSC(p)$ with noiseless feedback, i.e.
\begin{equation}\label{deferexp}
F(R,p)=\limsup_{n\to\infty}\frac{1}{n} \ln\frac{1}{P_{\rm e}(M_n,n,p)},\quad
M_n=e^{Rn},
\end{equation}
where $P_{\rm e}(M_n,n,p)$ is defined in \eqref{deferprob}. Clearly, the
function $F(R,p)$ is non-increasing in $R$.

Introduce also the limiting value $F(0,p)$
\begin{equation}\label{deferexp1a}
F(0,p)=\lim_{R\to 0} F(R,p),\quad 0<p<1/2.
\end{equation}
The limit in \eqref{deferexp1a} is well-defined, since the
function $F(R,p)$ is bounded and non-increasing in $R$.

Equivalently, the function $F(0,p)$ is defined by \eqref{deferexp}, if the
number of messages $M_n$ is such that $M_n\to\infty$, but
$\log M_n = o(n)$ as $n\to\infty$.

Similarly, define by $F_K(p)$, $K=2,3,\ldots\strut$, the best error exponent for $K$ codewords over $\rm BSC(p)$ with noiseless feedback, i.e.
\begin{equation}\label{deferexp1}
F_K(p)=\limsup_{n\to\infty}\frac{1}{n} \ln \frac{1}{P_{\rm e}(K,n,p)},
\end{equation}
where $P_{\rm e}(K,n,p)$ -- minimal possible decoding error
probability (for all transmission methods of length $n$).
It was shown in the paper \cite{Ber1}, that
\begin{equation}\label{deferexp2}
F_3(p)=F_4(p)=\ldots=F(0,p),
\end{equation}
and therefore for investigation of the function $F(0,p)$ it is sufficient to find the value~$F_3(p)$.

Denote by $E_k(p)$, $k\ge 2$, the best error exponent for $k$ codewords
over $\rm BSC(p)$. Clearly,
$$
E_2(p)=F_2(p)=\frac{1}{2}\ln\frac{1}{4pq}.
$$
It is also clear that $E_3(p)$ is defined by $n$-simplex code
$(\bfm{x}_1,\bfm{x}_2,\bfm{x}_3)$ (i.e. the code, for which
$d(\bfm{x}_i,\bfm{x}_j)\approx 2n/3$ for all $i\ne j$), and therefore
$$
E_3(p)=\frac{1}{3}\ln\frac{1}{4pq}.
$$
Clearly, we have $E_3(p)\le F_3(p)\le E_2(p)$.

The next result was proved in \cite{Ber1}.

{\bf Proposition.} {\it For $P_{\rm e}(3,n,p)$ the upper bound holds}
({\it see }\eqref{deferexp1}, \eqref{deferexp2})
\begin{equation}\label{prop1}
P_{\rm e}(3,n,p)\le\Bigl(\frac{q}{p}\Bigr)^{1/3}
\bigl(p^{1/3}q^{2/3}+p^{2/3}q^{1/3}\bigr)^n.
\end{equation}

It follows from \eqref{deferexp1a}, \eqref{deferexp2} and \eqref{prop1} that
\begin{equation}\label{F0p}
F_3(p)\ge F_{\rm fb}(p),
\end{equation}
where
\begin{equation}\label{Ffp}
F_{\rm fb}(p)=-\ln\bigl(p^{1/3}q^{2/3}+p^{2/3}q^{1/3}\bigr)
=-\ln\bigl[p^{1/3}q^{2/3}(1+z^{1/3})\bigr]\ge 0
\end{equation}
and
$$
p+q=1,\qquad z=z(p)=p/q.
$$
In particular, the formula \eqref{F0p} followed also from subsequent papers
\cite{Zig1,Bur88} (where by other methods the whole reliability function $F(R)$ was investigated).

Moreover, it was also claimed in \cite{Ber1} that the opposite to \eqref{F0p} inequality holds
\begin{equation}\label{F0p1}
F_3(p)\le F_{\rm fb}(p),
\end{equation}
and if the formula \eqref{F0p1} is correct, then from \eqref{F0p} the equality would follow
\begin{equation}\label{F0p2}
F_3(p)=F_{\rm fb}(p).
\end{equation}

However, there was no rigorous proof of formula \eqref{F0p1} in \cite{Ber1}. Later, in \cite{Zig78} one more attempt to establish formula \eqref{F0p1} was done (using general Bellman's equation), but later it was found that the proof is also incorrect.

Below in the paper formula \eqref{F0p1} is proved, and therefore formula
\eqref{F0p2} holds.

We describe all possible transmission methods of one of three messages over
$\rm BSC$ with noiseless feedback. Note that any reasonable transmission strategy
has the following form. At each instant $k$, $k=1,\ldots,n$, based on previous channel outputs $\bfm{y}^{k-1}$, the receiver selects some message $\theta_{i_0}$, and asks the transmitter whether $\theta_{i_0}$ is the true message $\theta_{\rm true}$. Here the noiseless feedback is important !
If the true message $\theta_{\rm true}$ coincides with~~$\theta_{i_0}$, i.e. $\theta_{\rm true}=\theta_{i_0}$, then the signal $x_k=0$ is transmitted. If
$\theta_{\rm true}\ne\theta_{i_0}$, then the signal $x_k=1$ is transmitted.
After the instant $n$ the decision is made in favor of the most probable message $\theta_i$.

The transmission strategy used in papers \cite{Ber1,Zig1,Bur88} is quite
natural: at each instant $k$ as $\theta_{i_0}(k)$ the most probable message
$\theta_i$, provided output $\bfm{y}^{k-1}$ is selected. It seems that
such transmission strategy provides the best decoding error exponent $F_3(p)$
(that statement needs to be proved, what was not done in \cite{Ber1,Zig1,Bur88}).

The main result of the paper is as follows.

{\bf Theorem 1.}
{\it For $P_{\rm e}(3,n,p)$ the lower bound is valid
\begin{equation}\label{deferprob1}
P_{\rm e}(3,n,p)\ge\frac{1}{2}\bigl(p^{1/3}q^{2/3}+p^{2/3}q^{1/3}\bigr)^n
=\frac{1}{2}\bigl[p^{1/3}q^{2/3}(1+z^{1/3})\bigr]^n,\quad z=p/q,
\end{equation}
and therefore for $F_3(p)$ the formula holds} ({\it see} \eqref{Ffp})
\begin{equation}\label{lowF3p}
F_3(p)= F(0,p)=F_{\rm fb}(p).
\end{equation}

Note that,
\begin{equation}\label{lowF23p}
E_2(p)=F_2(p)>F_3(p)>E_3(p),\quad 0<p<1/2.
\end{equation}

For $\rm BSC$ output denote $\bfm{y}^k=y_1^k=(y_1,\ldots,y_k)$, $k=1,\ldots,n$, $y_k\in\{0,1\}$.

{\it Remark 1}.
Let us explain why for three messages noiseless feedback can
help to improve the decoding error probability. Indeed, assume that at a time
instant $i$ we have
$$
p(\bfm{y}^i| \bfm{x}_1) \approx p(\bfm{y}^i| \bfm{x}_2 ) \gg p(\bfm{y}^i|
\bfm{x}_3),
$$
i.e., the message $\theta_3$ is much less probable than messages
$\theta_1,\theta_2$ (and due to noiseless feedback it is known at the
transmitter !). Then for time instants $t>i$ we may mainly test remaining
messages $\theta_1$ and $\theta_2$ (for example, using for that purpose opposite
code blocks, as for two messages). Since $E_2(p)>E_3(p)$ (see \eqref{lowF23p}),
then such coding would allow to decrease the decoding error probability.

{\it Remark 2}.
The right-hand side of the formula \eqref{Ffp} has the following useful
interpretation (not recorded earlier). Let
$(\bfm{x}_1,\bfm{x}_2,\bfm{x}_3)$ -- $n$-simplex code (i.e., the code,
such that  ~$d(\bfm{x}_i,\bfm{x}_j)\approx 2n/3$ for all $i\ne j$). Then the
formula holds
(see ~proof in Appendix)
\begin{equation}\label{property3}
\P\{\mathcal{E}_n\} \sim e^{-F_{\rm fb}(p)n},
\end{equation}
where
\begin{equation}\label{property3c}
\mathcal{E}_n=\bigl\{\bfm{y}^n:\: p(\bfm{y}^n| \theta_1) \approx p(\bfm{y}^n|
\theta_2) \approx p(\bfm{y}^n| \theta_3)\bigr\}.
\end{equation}
In other words, the event $\mathcal{E}_n$ defines the exponent of the
decoding error probability.

{\it Remark 3}.
From the viewpoint of reliability functions behavior the channel $\rm BSC(p)$
and Gaussian channel ~$G(A)$ with limitation on average power $A$ are mainly
similar to each other \cite{SH1, G1, Pin1}. But the same channels with noiseless
feedback show also an essential difference. In particular, for $\rm BSC(p)$ we
have $F_3(p)<E_2(p)$ (see ~\eqref{lowF23p}), while for Gaussian channel ~$G(A)$
we have $F_3(A)=E_2(A)$ (see \cite{Pin1}). That difference is based on the
feature that for Gaussian channel ~ $G(A)$ at some instants it is possible to
transmit
very strong signals, while it is impossible for the channel $\rm BSC(p)$.

Next result describes the best transmission method in the case of three messages.

{\bf Theorem 2.} {\it At each instant $k$, $k=1,\ldots,n$, the best partition
of messages $\{\theta_1,\theta_2,\theta_3\}$} ({\it minimizing the decoding error
probability $\P_{\rm e}(n)$}) {\it has the form: the most probable message}
({\it provided output $\bfm{y}^{k-1}$}) {\it versus two remaining messages.}

The paper is organized as follows. For a completeness purpose, in \S\,2 a short and elegant proof of formula \eqref{prop1} from \cite{Ber1} is presented (it seems that such proof is available only in the thesis \cite{Ber1} and it was not published in other more available sources). In \S\,3 Theorem 2 is proved. In
\S\,4 the Markov diagram for the decoder of the optimal transmission strategy
is introduced and described. In \S\,5 using that diagram Theorem 1 is proved.

\section{Proof of Proposition}
By $d(\bfm{y},\bfm{x})$ we denote the Hamming distance between vectors
$\bfm{y}$ and $\bfm{x}$. For each instant $k$, $k=1,\ldots,n$, by
$\theta^{(1)}(k),\theta^{(2)}(k),\theta^{(3)}(k)$ we denote the ordering of messages $\theta_1,\theta_2,\theta_3$ provided $\bfm{y}^k$, such that
\begin{equation}\label{order1}
p\bigl(\bfm{y}^k| \theta^{(1)}(k)\bigr)\ge p\bigl(\bfm{y}^k|
\theta^{(2)}(k)\bigr)\ge p\bigl(\bfm{y}^k| \theta^{(3)}(k)\bigr).
\end{equation}
By $\bfm{x}^{(1)}(k),\bfm{x}^{(2)}(k),\bfm{x}^{(3)}(k)$ we denote the corresponding ordering of codewords used. Then \eqref{order1} is equivalent to the ordering
$$
d\bigl(\bfm{y}^k,\bfm{x}^{(1)}(k)\bigr)\le d\bigl(\bfm{y}^k,\bfm{x}^{(2)}(k)\bigr)\le
d\bigl(\bfm{y}^k,\bfm{x}^{(3)}(k)\bigr).
$$
We call $d^{(i)}(k)=d(\bfm{y}^k,\bfm{x}^{(i)}(k))$ as the number of ``negative votes'' against $\theta^{(i)}(k)$ during the time $k$. Denote also $d_i=d_i(n)$.

Denote by $d_{1,3}(k)$, $k=1,\ldots,n$, the average number of
``negative votes'' against all messages during the time $k$, i.e.,
\begin{equation}\label{order2a}
d_{1,3}(k)=\frac{1}{3}\sum_{i=1}^{3}d^{(i)}(k).
\end{equation}

We use the strategy, when at each instant $k$ the most probable message $\theta^{(1)}(k)$ is selected, and the transmitter answers whether $\theta^{(1)}(k)$ is the true message $\theta_{\rm true}$. If
$\theta_{\rm true}=\theta^{(1)}(k)$, then the transmitter sends the signal  $x_k=0$, while if $\theta_{\rm true}\ne\theta^{(1)}(k)$, then the signal $x_k=1$ is sent.

Therefore, if the output signal $y_k=1$, then the message $\theta^{(1)}(k)$ gets one additional negative vote, while remaining two messages $\{\theta^{(2)}(k),\theta^{(3)}(k)\}$ do not get additional negative votes. If the output signal $y_k=0$, then the message $\theta^{(1)}(k)$ does not get additional negative votes, while each of remaining messages $\theta^{(2)}(k)$ and $\theta^{(3)}(k)$ gets one additional negative vote. As a result, if $y_k=1$, then the value $d_{1,3}$ from \eqref{order2a} increases by $1/3$.
If $y_k=0$, then the value $d_{1,3}$ increases by $2/3$. If $m$ zeros and $n-m$ ones were received on the output during the total time ~$n$, then
$d_{1,3}(n)=(n+m)/3$. There are $\nbinom{n}{m}$ ways to set $m$ zeros on $n$ positions.

For each instant $k$ the following inequalities hold
\begin{equation}\label{order2b}
d^{(1)}(k)\le d^{(2)}(k)\le d^{(3)}(k)\le d^{(2)}(k)+1.
\end{equation}
Only the last one of inequalities \eqref{order2b} should be explained. Indeed, it is true for $k=1$ (i.e., after getting an output $y_1$). Further, for
$k\ge 2$ for the strategy used messages $\theta^{(2)}(k)$ and $\theta^{(3)}(k)$ always fall in one group, and therefore the condition
$d^{(3)}(k)\le d^{(2)}(k)+1$ remains valid (although messages $\theta^{(2)}(k)$ and $\theta^{(3)}(k)$ themselves may change).

From \eqref{order2a} and \eqref{order2b} the inequality follows
\begin{equation}\label{order2c2}
d^{(2)}(k)\ge d_{1,3}(k)-1/3.
\end{equation}

Note that each realization of an output $\bfm{y}^n$ with $e$ errors has the probability $p^{e}q^{n-e}$. Since the true message gets $e$ negative votes, then for decoding error it is necessary to have $d^{(2)}(n)=e$ or $d^{(3)}(n)=e$.
In either case, by \eqref{order2c2} we need $e\ge d_{1,3}(n)-1/3$, and therefore it is necessary to have
\begin{equation}\label{order2d1}
p^{e}q^{n-e}\le\Bigl(\frac{q}{p}\Bigr)^{1/3}p^{d_{1,3}(n)}q^{n-d_{1,3}(n)}.
\end{equation}
Condition \eqref{order2d1} bounds the probability of any erroneous path
via the value $d_{1,3}(n)$. Note that if $m$ -- the number of zeros, received at output during all time~$n$, then the value $d_{1,3}(n)=(n+m)/3$, $m=0,1,\ldots,n$ corresponds to every erroneous path. Since there are
$\nbinom{n}{m}$ ways to distribute $m$ zeros on~$n$~positions, then by \eqref{order2d1} we get
$$
P_{\rm e}(3,n,p)\le\Bigl(\frac{q}{p}\Bigr)^{1/3}\sum_{m=0}^n\binom{n}{m}
p^{(n+m)/3}q^{(2n-m)/3}=\Bigl(\frac{q}{p}\Bigr)^{1/3}
\bigl(p^{1/3}q^{2/3}+p^{2/3}q^{1/3}\bigr)^n,
$$
from where \eqref{prop1} follows. $\triangle$

\section{Proof of Theorem 2}
We consider transmission of three equiprobable messages
$\{\theta_1,\theta_2,\theta_3\}$. After each instant $k$ we find
posterior message probabilities $\pi_i(k)$, $i=1,2,3$, based on
received block $\bfm{y}^k=y_1^k=(y_1,\ldots,y_k)$, $k=1,\ldots,n$.
Transmission at instant $k+1$ depends only on probabilities $\{\pi_i(k)\}$
(since they constitute a sufficient statistics). We may assume that at instant $k+1$ we start transmission, but using prior probabilities $\{\pi_i(k)\}$.

We denote by $d_i(k)=d_i(\bfm{y}^k)=d(\bfm{y}^k,\bfm{x}_i(k))$ the total number of ``negative votes'' against $\theta_i$ during the time $[1,k]$. Denote also $d_i=d_i(n)$.

All information the decoder has at an instant $k$, $k=1,\ldots,n$ after
receiving an output $\bfm{y}^k$, are posterior probabilities $\pi_i(\bfm{y}^k)$ of messages~$\theta_i$, $i=1,2,3$ (or, equivalently, the set of distances $d_i(\bfm{y}^k)$, $i=1,2,3$). Denote by $i_0(\bfm{y}^k)\in\{1,2,3\}$, the index providing the maximal value to $\pi_i(\bfm{y}^k)$ (or, equivalently, the minimal value to $d_i(\bfm{y}^k))$, i.e.
\begin{equation}\label{oneinst80}
\pi_{i_0(\bfm{y}^k)}(\bfm{y}^k)=\max_i\pi_i(\bfm{y}^k),\quad
d_{i_0(\bfm{y}^k)}(\bfm{y}^k)=\min_id_i(\bfm{y}^k),\quad k=1,\ldots,n.
\end{equation}
From the decoder viewpoint the value $\pi_i(\bfm{y}^n)$ is the posterior probability of the event $\{\theta_i=\theta_{\rm true}\}$. Therefore, best (from the decoding error probability viewpoint) is to make the decision in favor of the message $\theta_{i_0(\bfm{y}^n)}$ with the maximal posterior probability $\pi_{i_0(\bfm{y}^n)}(\bfm{y}^n)$. Then we have by \eqref{oneinst80}
\begin{equation}\label{oneinst8}
P_{\rm e}(n)=\P\{\theta_{i_0(\bfm{y}^n)}\ne\theta_{\rm true}\}=1-\E
I_{\{\theta_{i_0(\bfm{y}^n)}=\theta_{\rm true}\}}=1-\E\pi_{i_0(\bfm{y}^n)}.
\end{equation}

By $\mathcal{A}_k\in\{\theta_1, \theta_2, \theta_3\}$, $k=1,\ldots,n$, we denote the message, selected by the receiver at instant $k$, on which it asks the question, whether $\mathcal{A}_k$ is the true message $\theta_{\rm true}$.

Consider changing of the value $\pi_{i_0(\bfm{y}^k)}$ from \eqref{oneinst80}, \eqref{oneinst8} depending on a choice of the message $\mathcal{A}_{k+1}$.
For that purpose it is sufficient to consider changing of the value
$$
\sum\limits_{j\ne i_0}z^{d_j(k)-d_{i_0}(k)},
$$
where $z= p/q$, i.e., changing of the value $1/\pi_{i_0(\bfm{y}^k)}$ (see formulas \eqref{order3}, \eqref{order31}).

Two cases are possible:
\begin{enumerate}[1.]
\item\vskip-5pt
There exists a unique index $i_0(\bfm{y}^k)$, such that
$d_j(\bfm{y}^k)- d_{i_0(\bfm{y}^k)}(\bfm{y}^k)\ge 1$ for all
$j\ne i_0(\bfm{y}^k)$. Then $i_0(\bfm{y}^{k+1})=i_0(\bfm{y}^k)$ for all $y_{k+1}$. In that case the most probable message $\theta_{i_0(\bfm{y}^k)}$ at instant $k$ remains the same for instant $k+1$ for any output $y_{k+1}$.
\item\vskip1pt
There are two different indices $i_0(\bfm{y}^k)$ and $i_1(\bfm{y}^k)$,
such that
$d_{i_0(\bfm{y}^k)}(\bfm{y}^k)=d_{i_1(\bfm{y}^k)}(\bfm{y}^k)$ and
$d_j(\bfm{y}^k)-d_{i_0(\bfm{y}^k)}(\bfm{y}^k)\ge 1$ for the third index.
\end{enumerate}\vskip-5pt

It is clear that in the third possible case (when all distances
$d_j(\bfm{y}^k)$, $j=1,2,3$, are equal) due to symmetry any choice of
$\mathcal{A}_{k+1}$ (i.e. any partition of messages) leads to the same result.

Consider first the case 1). Denote for short (where $z=p/q$)
\begin{equation}\label{oneinst8a3}
\begin{gathered}
a_j=a_j(\bfm{y}^k)=z^{d_j(\bfm{y}^k)-d_{i_0(\bfm{y}^k)}(\bfm{y}^k)},\\
\delta_j=\delta_j(y_{k+1})=d_j(y_{k+1})- d_{i_0(\bfm{y}^k)}(y_{k+1}),\\
B(k,\bfm{y}^k) =\sum\limits_{j\ne
i_0}z^{d_j(\bfm{y}^k)-d_{i_0(\bfm{y}^k)}(\bfm{y}^k)}=\sum\limits_{j\ne i_0}a_j,\\
B(k+1)=\sum\limits_{j\ne i_0}z^{d_j(\bfm{y}^{k+1})-
d_{i_0(\bfm{y}^{k+1})}(\bfm{y}^{k+1})}=\sum\limits_{j\ne i_0}z^{d_j(\bfm{y}^{k+1})-
d_{i_0(\bfm{y}^k)}(\bfm{y}^{k+1})}=\sum\limits_{j\ne i_0}a_jz^{\delta_j(y_{k+1})}.
\end{gathered}
\end{equation}
Denote also
\begin{equation}\label{oneinst8a31}
B_j(k+1)= B(k+1)\quad \text{if}\quad \mathcal{A}_{k+1}=\theta_j,\quad j=1,2,3.
\end{equation}
Note that values $\delta_j$, $j=1,2,3$, take on only values $0,1$ and $-1$.
Without loss of generality we may assume that $i_0(\bfm{y}^k) =1$, and therefore $i_0(\bfm{y}^{k+1})=1$. Then we have $\delta_1(y_{k+1})=0$ and
\begin{equation}\label{oneinst8a4}
\begin{aligned}
B(k)&= a_2+a_3,&\qquad B(k+1) &=a_2z^{\delta_2(y_{k+1})}+a_3z^{\delta_3(y_{k+1})},\\
\pi_1(k)&=\frac{1}{1+B(k)},&\qquad \pi_1(k+1)&=\frac{1}{1+B(k+1)}.
\end{aligned}
\end{equation}
Consider distributions of the random variables $B_j(k+1)$, $j=1,2,3$, provided
$i_0(\bfm{y}^k) =1$. For $j=1$, i.e. $\mathcal{A}_{k+1}=\theta_1$, we have
\begin{equation}\label{oneinst8a5}
\delta_2=\delta_3=
\begin{cases}
1 & \text{with probability}\ \pi_1(k)q+(1-\pi_1(k))p=p+(q-p)\pi_1(k),\\ -1 &
\text{with probability}\ q-(q-p)\pi_1(k),
\end{cases}
\end{equation}
and then
\begin{equation}\label{oneinst8a6}
B_1(k+1)=
\begin{cases}
(a_2+a_3)z=B(k)z & \text{with probability}\ p+(q-p)\pi_1(k),\\ (a_2+a_3)/z=B(k)/z &
\text{with probability}\ q- (q-p)\pi_1(k).
\end{cases}
\end{equation}
For $j=2$, i.e., $\mathcal{A}_{k+1}= \theta_2$, we have
\begin{equation}\label{oneinst8a7}
\delta_3=0,\quad \delta_2=
\begin{cases}
1 & \text{with probability}\ \pi_1(k)q+(1-\pi_1(k))p=p+(q-p)\pi_1(k),\\ -1 &
\text{with probability}\ q-(q-p)\pi_1(k),
\end{cases}
\end{equation}
and therefore
\begin{equation}\label{oneinst8a8}
B_2(k+1)=
\begin{cases}
a_2z+a_3 & \text{with probability}\ p+(q-p)\pi_1(k),\\ a_2/z+a_3 & \text{with
probability}\ q-(q-p)\pi_1(k).
\end{cases}
\end{equation}
Similarly, for $j=3$, i.e., $\mathcal{A}_{k+1}=\theta_3$, we have
\begin{equation}\label{oneinst8a9}
\delta_2=0,\quad \delta_3=
\begin{cases}
1 & \text{with probability}\ p+(q-p)\pi_1(k),\\ -1 & \text{with probability}\
q-(q-p)\pi_1(k),
\end{cases}
\end{equation}
and then
\begin{equation}\label{oneinst8a10}
B_3(k+1)=
\begin{cases}
a_3z+a_2 & \text{with probability}\ p+(q-p)\pi_1(k),\\ a_3/z+a_2 & \text{with
probability}\ q-(q-p)\pi_1(k).
\end{cases}
\end{equation}

As a result, we have for $i_0(\bfm{y}^k) =1$
\begin{equation}\label{order4}
\begin{aligned}[b]
E_1&=\E\bigl[\pi_{i_0}(k+1)| \bfm{y}^k,\mathcal{A}_{k+1}=\theta_1\bigr]
=\E\biggl[\frac{1}{1+ B_1(k+1)}\Bigm| \bfm{y}^k,\mathcal{A}_{k+1}=\theta_1\biggr]\\
&=\frac{p+(q-p)\pi_1(k)}{1+(a_2+a_3)z}+\frac{q-(q-p)\pi_1(k)}{1+(a_2+a_3)/z}.
\end{aligned}
\end{equation}
Similarly we have
\begin{equation}\label{order4a}
\begin{aligned}[b]&
E_2=\E\bigl[\pi_{i_0}(k+1)|
\bfm{y}^k,\mathcal{A}_{k+1}=\theta_2\bigr]=\E\biggl[\frac{1}{1+ B_2(k+1)}\Bigm|
\bfm{y}^k,\mathcal{A}_{k+1}=\theta_2\biggr]\\ &=
\frac{p+(q-p)\pi_1(k)}{1+a_2z+a_3}+\frac{q-(q-p)\pi_1(k)}{1+a_2/z+a_3}
\end{aligned}
\end{equation}
and
\begin{equation}\label{order4b}
\begin{aligned}[b]&
E_3=\E\bigl[\pi_{i_0}(k+1)| \bfm{y}^k,\mathcal{A}_{k+1}=\theta_3\bigr]
=\E\biggl[\frac{1}{1+ B_3(k+1)}\Bigm|\bfm{y}^k,\mathcal{A}_{k+1}=\theta_3\biggr] \\
&=\frac{p+(q-p)\pi_1(k)}{1+a_2+a_3z}+\frac{q-(q-p)\pi_1(k)}{1+a_2+a_3/z}.
\end{aligned}
\end{equation}

We shall show that $E_1\ge\max\{E_2,E_3\}$, what means that best is to
use $\mathcal{A}_{k+1}=\theta_1=\theta_{i_0(\bfm{y}^k)}$.
Due to symmetry it is sufficient to show that $E_1\ge E_2$. Indeed,
by \eqref{order4} and \eqref{order4a} we have
\begin{equation}\label{order4c}
\begin{aligned}[b]
E_1-E_2&= [p+(q-p)\pi_1(k)]\left[\frac{1}{1+(a_2+a_3)z}-\frac{1}{1+a_2z+a_3}\right]\\
&\quad\strut+[q-(q-p)\pi_1(k)]\left[\frac{1}{1+(a_2+a_3)/z}-
\frac{1}{1+a_2/z+a_3}\right]\\ &=
\frac{[p+(q-p)\pi_1(k)]a_3(1-z)}{[1+(a_2+a_3)z][1+a_2z+a_3]}
+\frac{[q-(q-p)\pi_1(k)]a_3(1-1/z)}{[1+(a_2+a_3)/z][1+a_2/z+a_3]}\\ &=
a_3(1-z)\biggl\{\frac{p+(q-p)\pi_1(k)}{[1+(a_2+a_3)z][1+a_2z+a_3]}
-\frac{q-(q-p)\pi_1(k)}{z[1+(a_2+a_3)/z][1+a_2/z+a_3]}\biggr\}\\
&=qa_3(1-z)\left\{\frac{z+(1-z)\pi_1(k)}{[1+(a_2+a_3)z](1+a_2z+a_3)}
-\frac{1-(1-z)\pi_1(k)}{(z+a_2+a_3)(1+a_2/z+a_3)}\right\}.
\end{aligned}
\end{equation}
It is sufficient to show, that
$$
\frac{z+(1-z)\pi_1(k)}{[1+(a_2+a_3)z](1+a_2z+a_3)}
-\frac{1-(1-z)\pi_1(k)}{(z+a_2+a_3)(1+a_2/z+a_3)}\ge 0,
$$
or, equivalently (after a standard algebra using the formula
$\pi_1(k)=1/(1+a_2+a_3)$),
\begin{equation}\label{order4d}
a_2(1-z)\ge 0.
\end{equation}
The relation \eqref{order4d} holds, if $z\le 1$ (i.e., if $p\le 1/2$).
By \eqref{order4c} and \eqref{order4d} we have $E_1\ge E_2$.
Similarly we get $E_1\ge E_3$. Therefore, $E_1\ge\max\{E_2,E_3\}$, what means that best is to use
$\mathcal{A}_{k+1}=\theta_1=\theta_{i_0(\bfm{y}^k)}$.
It completes considering of the case 1).

Consider now the case 2), when there are two different indices
$i_0(\bfm{y}^k)$ and $i_1(\bfm{y}^k)$, such that
$d_{i_0(\bfm{y}^k)}(\bfm{y}^k)=d_{i_1(\bfm{y}^k)}(\bfm{y}^k)$ and
$d_j(\bfm{y}^k)-d_{i_0(\bfm{y}^k)}(\bfm{y}^k)\ge 1$
for the third index. Without loss of generality we may assume that
$i_0(\bfm{y}^k) =1$ and $i_1(\bfm{y}^k) =3$. Then $E_1=E_3$,
and it remains to show that $E_1\ge E_2$, and then best is to use
$\mathcal{A}_{k+1}=\theta_1=\theta_{i_0(\bfm{y}^k)}$
(or $\mathcal{A}_{k+1}=\theta_3=\theta_{i_1(\bfm{y}^k)}$).
Note that for any ~$y_{k+1}$ one of distances
$d_{i_0(\bfm{y}^{k+1})}(\bfm{y}^{k+1})$ or
$d_{i_1(\bfm{y}^{k+1})}(\bfm{y}^{k+1})$ remains the same as earlier for the
instant $k$. Remaining calculations essentially coincide with
\eqref{oneinst8a3}-\eqref{order4d} (~in fact, they are even simpler) and we
omit them. It completes the proof of Theorem~2. $\triangle$

\section{Markov diagram of the optimal strategy decoder }
Introduce the Markov chain describing the decoder evolution in time. Denote by
$d_i(k)=d(\bfm{y}^k,\bfm{x}_i(k))$ the total number of ``negative votes'' against~$\theta_i$ during the transmission period $[1,k]$. Denote also
$d_i=d_i(n)$. Then ($z=p/q<1$)
\begin{equation}\label{order3}
\pi_i(k)=\frac{z^{d_i(k)}}{\sum\limits_{j=1}^{3}z^{d_j(k)}}=\frac{1}{1+\sum\limits_{j\ne
i}z^{d_j(k)-d_i(k)}},\qquad \pi_i(n)=\frac{1}{1+\sum\limits_{j\ne
i}z^{d_j(n)-d_i(n)}}.
\end{equation}
Note that,
\begin{equation}\label{order31}
\frac{\pi_i(k)}{1-\pi_i(k)}=\frac{z^{d_i(k)}}{\sum\limits_{j\ne
i}z^{d_j(k)}}=\frac{1}{\sum\limits_{j\ne i}z^{d_j(k)-d_i(k)}}.
\end{equation}

For each instant $k$ and each output $\bfm{y}^k$ define for a message $\theta_i$  the metrics $m_i(k,\bfm{y}^k)$ as follows:
\begin{equation}\label{mardef1}
m_i(k,\bfm{y}^k)=d\bigl(\bfm{y}^k,\bfm{x}_i(k)\bigr)-\min_j
d\bigl(\bfm{y}^k,\bfm{x}_j(k)\bigr)=d_i(k)-\min_jd_j(k),\quad i=1,2,3.
\end{equation}
Clearly, $m_i(k,\bfm{y}^k)\ge 0$ and $\min\limits_im_i(k,\bfm{y}^k)=0$.
The set $\{m_i(k,\bfm{y}^k)\}$ is a sufficient statistics, since it defines
posterior probabilities $\{\pi_i(k)\}$ (see~\eqref{order3}--\eqref{mardef1}).

Denote by $S_{ij\ell}=S_{ijl}(k)=S_{ijl}(k,\bfm{y}^k)$ the chain state
with $i=m_1(k,\bfm{y}^k)$, $j=m_2(k,\bfm{y}^k)$, $\ell=m_3(k,\bfm{y}^k)$.

As a result, the whole diagram looks like an ``octopus'' with nine ``tentacles'' (see Fig.~1). For example, one of such ``tentacles'' is
$(S_{011},S_{022},S_{033},\ldots)$.

\begin{figure}[tp]
\unitlength=0.00125\textwidth \centering
\begin{picture}(360,340)(-10,-80)
\thicklines \put(190,80){\circle*{6}} \put(60,80){\circle*{6}}
\put(0,80){\circle*{6}} \put(320,80){\circle*{6}} \put(347,40){\circle*{6}}
\put(347,120){\circle*{6}} \put(125,180){\circle*{6}} \put(255,180){\circle*{6}}
\put(282,220){\circle*{6}} \put(125,-20){\circle*{6}} \put(152,220){\circle*{6}}
\put(98,220){\circle*{6}} \put(255,-20){\circle*{6}} \put(282,-60){\circle*{6}}
\put(152,-60){\circle*{6}} \put(98,-60){\circle*{6}} \put(190,80){\vector(-1,0){130}}
\put(190,80){\vector(1,0){130}} \put(190,80){\vector(-2,3){65}}
\put(190,80){\vector(2,3){65}} \put(190,80){\vector(2,-3){65}}
\put(190,80){\vector(-2,-3){65}} \put(125,180){\vector(1,0){130}}
\put(125,-20){\vector(1,0){130}} \put(200,85){000} \put(35,87){011}
\put(330,75){\!100} \put(100,176){\!\!001} \put(260,176){101} \put(95,-24){010}
\put(260,-24){110} \put(-10,87){022} \put(290,212){202} \put(290,-65){220}
\put(160,-65){120} \put(108,-65){021} \put(355,38){210} \put(160,212){102}
\put(107,217){012} \put(355,115){201} \put(125,180){\vector(-2,-3){65}}
\put(320,80){\vector(-2,-3){65}} \put(320,80){\vector(-2,3){65}}
\put(125,-20){\vector(-2,3){65}} \put(60,80){\vector(-1,0){50}}
\put(0,80){\vector(1,0){50}} \put(0,80){\vector(-1,0){50}}
\put(-50,80){\vector(1,0){40}} \put(255,180){\vector(2,3){26}}
\put(282,220){\vector(2,3){20}} \put(282,220){\vector(-2,-3){20}}
\put(293,237){\vector(-2,-3){10}} \put(255,-20){\vector(2,-3){26}}
\put(282,-60){\vector(2,-3){20}} \put(282,-60){\vector(-2,3){20}}
\put(296,-82){\vector(-2,3){10}} \put(125,-20){\vector(2,-3){26}}
\put(152,-60){\vector(2,-3){20}} \put(152,-60){\vector(-2,3){20}}
\put(166,-82){\vector(-2,3){10}} \put(125,-20){\vector(-2,-3){26}}
\put(99,-60){\vector(-2,-3){20}} \put(98,-60){\vector(2,3){20}}
\put(85,-82){\vector(2,3){10}} \put(320,80){\vector(2,-3){26}}
\put(347,40){\vector(2,-3){20}} \put(347,40){\vector(-2,3){20}}
\put(361,20){\vector(-2,3){10}} \put(125,180){\vector(2,3){26}}
\put(152,220){\vector(2,3){20}} \put(152,220){\vector(-2,-3){20}}
\put(163,237){\vector(-2,-3){10}} \put(125,180){\vector(-2,3){26}}
\put(98,220){\vector(-2,3){20}} \put(98,220){\vector(2,-3){20}}
\put(86,237){\vector(2,-3){10}} \put(320,80){\vector(2,3){26}}
\put(347,120){\vector(2,3){20}} \put(347,120){\vector(-2,-3){20}}
\put(358,137){\vector(-2,-3){10}} \qbezier(60,80)(135,110)(190,80)
\put(120,95){\vector(1,0){15}} \qbezier(255,180)(200,155)(190,80)
\put(215,147){\vector(-2,-3){10}} \qbezier(255,-20)(240,60)(190,80)
\put(237,35){\vector(-2,3){10}}
\end{picture}\vskip5pt
\caption{ }
\end{figure}

We call $S_{000}$ the \emph{\/main}
state and six states $\{S_{011},S_{100},S_{101},S_{010},S_{110},S_{001}\}$ --
\emph{\/basic} states. Remaining states are on ``tentacles''.

For the decoding error probability $P_{\rm e}(n)$ we have
\begin{equation}\label{order2c}
P_{\rm e}(n)\ge\frac{2}{3}P_0(n),
\end{equation}
where
\begin{equation}\label{order2c1}
P_0(n)=P\{S_{000}(0) \Rightarrow S_{000}(n)\}.
\end{equation}

We describe transitions among states for the optimal strategy. Without loss of generality we may assume that $\theta_{\rm true}=\theta_1$.

If at instant $k$ the decoder is in the state $S_{000}(k)$,
then the set $\mathcal{A}(k+1)$ is chosen equiprobably among three possible
variants. As a result, for next possible state $S(k+1)$ we get
\begin{equation}\label{order2d}
S_{000}(k)\to
\begin{cases}
S_{011}(k+1) & \text{with probability}\ q/3,\\ S_{100}(k+1) & \text{with
probability}\ p/3,\\ S_{101}(k+1) & \text{with probability}\ p/3,\\ S_{010}(k+1) &
\text{with probability}\ q/3,\\ S_{110}(k+1) & \text{with probability}\ p/3,\\
S_{001}(k+1) & \text{with probability}\ q/3.
\end{cases}
\end{equation}
Indeed, at an instant $k$ each message $\theta_i$ has the probability
$\pi_i(k)=1/3$, $i=1,2,3$. Therefore with probability $1/3$
we have $\mathcal{A}(k+1)=\theta_1$. Since we assumed
$\theta_{\rm true}=\theta_1$, then with probability $q/3$ we get
$S(k+1)=S_{011}(k+1)$ and with probability $p/3$ we get $S(k+1)=S_{100}(k+1)$. Similarly remaining lines of \eqref{order2d} are obtained.

The easiest case is to describe transitions from states, for which the set
${\mathcal A}(k+1)$ is defined uniquely, without randomization (i.e., when
there is only one most probable message). Such states are
$S_{011}(k),S_{101}(k),S_{110}(k),\ldots\strut$. For those states we get
\begin{align}
S_{011}(k)\to
\begin{cases}
S_{000}(k+1) & \text{with probability}\ p,\\ S_{022}(k+1) & \text{with probability}\
q,
\end{cases}\label{order3d}\\
S_{101}(k)\to
\begin{cases}
S_{000}(k+1) & \text{with probability}\ q,\\ S_{202}(k+1) & \text{with probability}\
p
\end{cases}\label{order5d}\\
S_{110}(k)\to
\begin{cases}
S_{000}(k+1) & \text{with probability}\ q,\\ S_{220}(k+1) & \text{with probability}\
p.
\end{cases}\label{order7d}
\end{align}
Similarly, transitions from analogous states
$S_{022}(k),S_{202}(k),S_{220}(k),\ldots\strut$ are described. Transitions
from remaining states $S_{100}(k),S_{010}(k),S_{001}(k)$ are described similarly to \eqref{order2d}:
\begin{align}
S_{100}(k)\to
\begin{cases}
S_{101}(k+1) & \text{with probability}\ q/2,\\ S_{210}(k+1) & \text{with
probability}\ p/2,\\ S_{110}(k+1) & \text{with probability}\ q/2,\\ S_{201}(k+1) &
\text{with probability}\ p/2,
\end{cases}\label{order4d1}\\
S_{010}(k)\to
\begin{cases}
S_{021}(k+1) & \text{with probability}\ q/2,\\ S_{110}(k+1) & \text{with
probability}\ p/2,\\ S_{120}(k+1) & \text{with probability}\ p/2,\\ S_{011}(k+1) &
\text{with probability}\ q/2,
\end{cases}\label{order6d}\\
S_{001}(k)\to
\begin{cases}
S_{012}(k+1) & \text{with probability}\ q/2,\\ S_{101}(k+1) & \text{with
probability}\ p/2,\\ S_{102}(k+1) & \text{with probability}\ p/2,\\ S_{011}(k+1) &
\text{with probability}\ q/2.
\end{cases}\label{order8d}
\end{align}

\section{Proof of Theorem 1}
By \eqref{order2c}, \eqref{order2c1} it is sufficient to estimate from below
the value $P_0(n)=P\{S_{000}(0) \Rightarrow S_{000}(n)\}$. Clearly,
\begin{equation}\label{tot0}
P_0(n)=\sum_{t_n} \P\{t_n\},
\end{equation}
where the sum is taken over all paths $t_n$ of length $n$ and of the form
$S_{000}(0) \Rightarrow S_{000}(n)$.

We call $3$-path any path of length $3$ and of the form
$S_{000}(k) \Rightarrow S_{000}(k+3)$. We call also $2$-path any path of length $2$ and of the form $S_{000}(k) \Rightarrow S_{000}(k+2)$.

First, we limit ourselves in the right-hand side of \eqref{tot0} to paths $t_n$, passing only through the main and basic states (i.e., they do not pass through  tentacles). It is simple to see that any such path $t_n$ consists of
$3$-paths and $2$-paths.

There are six $3$-paths:
$$
\begin{gathered}
S_{000}\to S_{100}\to S_{101}\to S_{000}\quad \text{with probability}\ pq^2/6,\\
S_{000}\to S_{100}\to S_{110}\to S_{000}\quad \text{with probability}\ pq^2/6,\\
S_{000}\to S_{010}\to S_{011}\to S_{000}\quad \text{with probability}\ pq^2/6,\\
S_{000}\to S_{010}\to S_{110}\to S_{000}\quad \text{with probability}\ pq^2/6,\\
S_{000}\to S_{001}\to S_{101}\to S_{000}\quad \text{with probability}\ pq^2/6,\\
S_{000}\to S_{001}\to S_{011}\to S_{000}\quad \text{with probability}\ pq^2/6.
\end{gathered}
$$
Therefore
\begin{equation}\label{tot2}
P\{S_{000}(k)\to S_{000}(k+3)\}=pq^2.
\end{equation}

There are three $2$-paths:
\begin{equation}\label{tot1a}
\begin{gathered}
S_{000}\to S_{011}\to S_{000}\quad \text{with probability}\ qp/3,\\ S_{000}\to
S_{101}\to S_{000}\quad \text{with probability}\ qp/3,\\ S_{000}\to S_{110}\to
S_{000}\quad \text{with probability}\ qp/3.
\end{gathered}
\end{equation}
Therefore
\begin{equation}\label{tot1}
P\{S_{000}(k)\to S_{000}(k+2)\}=pq.
\end{equation}

We estimate the value $P_0(n)$ from \eqref{tot0}, using \eqref{tot2}--\eqref{tot1}. Any path $t_n$, limited to basic states, consists of some number $n_2$ of $2$-paths and some number $n_3$ of $3$-paths. Moreover,
$2n_2+3n_3= n$, $0\le n_2\le n/2$, and the total number of paths equals to
$$
m=n_2+n_3=\frac{n+n_2}3.
$$
There are $\nbinom{m}{n_2}$ ways to distribute $n_2$ \kern1pt$2$-paths.
Remaining $m-n_2$ places are occupied by~$n_3$ $3$-paths. Therefore we have ($z=p/q$)
\begin{equation}\label{tot51}
\begin{aligned}[b]
P_0(n)&=\sum_{n_2=0}^{n/2}\binom{(n+n_2)/3}{n_2}(pq)^{n_2} (pq^2)^{n_3}\\
&=\sum_{n_2=0}^{n/2}\binom{(n+n_2)/3}{n_2} (pq)^{n_2}(pq^2)^{(n-2n_2)/3}=(pq^2)^{n/3}
\sum_{n_2=0}^{n/2}\binom{(n+n_2)/3}{n_2}z^{n_2/3}.
\end{aligned}
\end{equation}

We estimate from below the sum in the right-hand side of \eqref{tot51}.
Maximum of the value $\nbinom{(n+n_2)/3}{n_2}z^{n_2/3}$ over $n_2$ is attained for $n_2=a_0n$, where the value $a_0$ will be found below. Then
\begin{equation}\label{tot51b}
\begin{aligned}[b]
\sum_{n_2=0}^{n/2}\binom{(n+n_2)/3}{n_2}z^{n_2/3}&\ge
\binom{n(1+a_0)/3}{a_0n}z^{a_0n/3}\\ &\ge\frac{1}{n}
\sum_{n_2=0}^{n(1+a_0)/3}\binom{n(1+a_0)/3}{n_2}z^{n_2/3}=
\frac{1}{n}(1+z^{1/3})^{n(1+a_0)/3}.
\end{aligned}
\end{equation}
In order to be accurate, we estimate also from above the sum in the right-hand side of \eqref{tot51}. We have
$$
\begin{aligned}
\sum_{n_2=0}^n\binom{(n+n_2)/3}{n_2}z^{n_2/3}&\le
n\binom{n(1+a_0)/3}{a_0n}z^{a_0n/3}\\ &\le
n\sum_{n_2=0}^{n(1+a_0)/3}\binom{n(1+a_0)/3}{n_2}z^{n_2/3}=
n(1+z^{1/3})^{n(1+a_0)/3}.
\end{aligned}
$$
As a result, we get from \eqref{tot51} and \eqref{tot51b}
\begin{equation}\label{tot51c}
P_0(n)\ge\frac{1}{n}(pq^2)^{n/3}(1+ z^{1/3})^{n(1+a_0)/3}.
\end{equation}
We find now the value $a_0$ in \eqref{tot51b}, \eqref{tot51c}. Since
$$
\ln\binom{(n+n_2)/3}{n_2} \approx \frac{(n+n_2)}{3}
h\biggl(\frac{3n_2}{n+n_2}\biggr),
$$
then denoting $n_2=an$, $0\le a\le 1/2$, introduce the function
$$
f_1(p,a)=(1+a)h\Bigl(\frac{3a}{1+a}\Bigr)-a\ln(q/p),\quad 0\le a\le 1/2.
$$

The value $a_0$ maximizes the function $f_1(p,a)$ over $0\le a\le 1/2$. Note that,
$$
\begin{gathered}
f_1(p,a)=(1+a)\ln(1+a)-3a\ln(3a)-(1-2a)\ln(1-2a)- a\ln(q/p),\\
(f_1(p,a))'_{a}=\ln\frac{p(1+a)(1-2a)^2}{27qa^{3}},\qquad (f_1(p,a))''_{aa}<0,\\
(f_1(p,a))'_{a=0}=\infty,\qquad (f_1(p,a))'_{a=1/2}=-\infty.
\end{gathered}
$$
Therefore, $a_0(p)$ is the unique root of the equation
$$
27qa^{3}-p(1+a)(1-2a)^2=0=(27-31p)a^{3} +3pa-p.
$$
For that root we have \cite[Ch.~1.8-3]{Korn}
$$
a_0(p)=\biggl[\frac{p}{2(27-31p)}\biggr]^{1/3}
\left\{\left[1+\sqrt{\frac{27(1-p)}{27-31p}}\,\right]^{1/3}
+\left[1-\sqrt{\frac{27(1-p)}{27-31p}}\,\right]^{1/3}\right\}.
$$

For small $p$ we have $3a_0(p) \approx p^{1/3}$. Since $a_0<2$, the estimate
\eqref{tot51c} yields to the upper bound \eqref{prop1}-\eqref{Ffp} for
$P_{\rm e}(3,n,p)$. But the estimate \eqref{tot51c} shows that when investigating the value $P_0(n)$, we may not limit ourselves only to basic states, but should take into account also states on tentacles.

We strengthen the estimate \eqref{tot51c}, taking also into account states on tentacles. We call by $2$-loop any path of length $2$ with the same starting and final states (not necessarily states $S_{000}$). Besides $2$-paths from  \eqref{tot1a}, other examples of $2$-loops are also
$$
\begin{aligned}&
S_{011}\to S_{022}\to S_{011}\quad \text{with probability}\ qp,\\ & S_{100}\to
S_{201}\to S_{100}\quad \text{with probability}\ qp/2,\\ & S_{100}\to S_{210}\to
S_{100}\quad \text{with probability}\ qp/2,\\ & S_{101}\to S_{202}\to S_{101}\quad
\text{with probability}\ qp,\quad \ldots.
\end{aligned}
$$
Such $2$-loops go out to tentacles.

We consider paths $t_n$, consisting of some number $n_3$ of $3$-paths and
some number $k_2$ of $2$-loops. Assume that we distributed $n_3$ $3$-paths
on $[1,n]$. After that we insert $k_2$ $2$-loops in any different $k_2$ instants on $[1,n]$. If such $2$-loop hits on the initial state of a $3$-path, then
that $3$-path is simply moved to the right on two steps. If such $2$-loop hits
an internal state of a $3$-path, then the part of that $3$-path is moved to the right on two steps, in order to imbed that $2$-loop. Similarly, $2$-loops can be inserted into other $2$-loops.

Since it is necessary to have $n=3n_3+2k_2$, then
\begin{equation}\label{sim1}
\begin{aligned}[b]
P_0(n)\ge\sum_{k_2=0}^{n/2}\binom{n}{k_2}(pq)^{k_2} (pq^2)^{n_3}
&=\sum_{k_2=0}^{n/2}\binom{n}{k_2}(pq)^{k_2}(pq^2)^{(n-2k_2)/3}\\
&=(pq^2)^{n/3}\sum_{k_2=0}^{n/2}\binom{n}{k_2}z^{k_2/3}.
\end{aligned}
\end{equation}
Note that,
$$
\binom{n}{k_2}z^{k_2/3}+\binom{n}{n-k_2}z^{(n-k_2)/3}\le 2\binom{n}{k_2}z^{k_2/3},
\quad k_2\le n/2,\quad z<1.
$$
Then \eqref{sim1} can be continued as follows:
\begin{equation}\label{sim3}
P_0(n)\ge\frac{1}{2}(pq^2)^{n/3}\sum_{k_2=0}^n\binom{n}{k_2}z^{k_2/3}
=\frac{1}{2}(pq^2)^{n/3}(1+z^{1/3})^n.
\end{equation}
Therefore from \eqref{sim3}, \eqref{order2c} and \eqref{order2c1} we get
\begin{equation}\label{order3c}
P_{\rm e}(n)\ge\frac{2}{3}P_0(n)\ge\frac{1}{3}(pq^2)^{n/3}(1+z^{1/3})^n.
\end{equation}
From \eqref{order3c} it follow \eqref{Ffp} and Theorem 1
(formulas \eqref{deferprob1}, \eqref{lowF3p}). $\triangle$

\hfill {\large\sl APPENDIX}

{\bf Proof of equation (\ref{property3}).}
Consider $n$-simplex code $(\bfm{x}_1,\bfm{x}_2,\bfm{x}_3)$, where
$\bfm{x}_1$ first has $n/3$ ones and then $2n/3$ zeros, $\bfm{x}_2$
first has $n/3$ zeros, then $n/3$ ones and then $n/3$ zeros, and $\bfm{x}_3$
first has $2n/3$ zeros and then $n/3$ ones. Then
$w(\bfm{x}_1)=w(\bfm{x}_2)=w(\bfm{x}_3)=n/3$ and
$d_{12}=d_{13}=d_{23}=2n/3$. Let an output $\bfm{y}$ has $u_1n/3$ ones on the
first $n/3$ positions, $u_2n/3$ ones on next $n/3$ positions and $u_3n/3$ ones
on last $n/3$ positions. Then
$$
\begin{gathered}
d(\bfm{x}_1,\bfm{y})/n=(1-u_1+u_2+u_3)/3,\\
d(\bfm{x}_2,\bfm{y})/n=(1+u_1-u_2+u_3)/3,\\
d(\bfm{x}_3,\bfm{y})/n=(1+u_1+u_2-u_3)/3.
\end{gathered}
$$
Since $d(\bfm{x}_1,\bfm{y})=d(\bfm{x}_2,\bfm{y})=d(\bfm{x}_3,\bfm{y})$, then we
get $u_1=u_2=u_3$ and
$$
p(\bfm{y}^n| \bfm{x}_1)=p^{d(\bfm{x}_1,\bfm{y})} q^{n-d(\bfm{x}_1,\bfm{y})}=q^n
z^{d(\bfm{x}_1,\bfm{y})}= q^nz^{(1+u)n/3},\quad z=p/q<1.
$$
Therefore
$$
\begin{aligned}
\P\bigl\{p(\bfm{y}^n| \bfm{x}_1) \approx p(\bfm{y}^n| \bfm{x}_2) \approx
p(\bfm{y}^n| \bfm{x}_3)\bigr\} &\sim \max_{0\le u\le 1}\P\Bigl\{p(\bfm{y}^n|
\bfm{x}_1) \approx q^nz^{(1+u)n/3}\Bigr\}\\ &\sim \max_{0\le u\le 1}
\Biggl\{\binom{n}{un}p^{(1+u)n/3}q^{(2-u)n/3}\Biggr\}\\ &\sim q^n\max_{0\le u\le
1}\Biggl\{\binom{n}{un}z^{(1+u)n/3}\Biggr\},
\end{aligned}
$$
and
\begin{equation}\label{property3a}
\frac{1}{n}\max_{0\le u\le 1}\ln\P\{p(\bfm{y}^n|
\bfm{x}_1)\}=\ln q+\max_{0\le
u\le 1}g(u),
\end{equation}
where
$$
g(u)=h(u)+(1+u)\ln(z^{1/3}),\qquad g'(u)=\ln\frac{1-u}{u}+\ln(z^{1/3}),\qquad
g''(u)<0.
$$
For the maximizing $u_0$ we get
$$
u_0 =\frac{1}{1+z^{-1/3}}= \frac{p^{1/3}}{p^{1/3}+ q^{1/3}},
$$
and after simple algebra
\begin{equation}\label{property3f}
\ln q+g(u_0)=\ln\bigl(p^{1/3}q^{2/3}+p^{2/3}q^{1/3}\bigr).
\end{equation}
From \eqref{property3a} and \eqref{property3f} formulas \eqref{property3} and
\eqref{property3c} follow. $\triangle$

%\section*{ACKNOWLEDGMENTS}
The author would like to thank Bassalygo L.A. and Kabatianski G.A. for
useful discussions and constructive critical remarks, which improved the paper.

%\section*{FUNDING}

%Supported in part by the Russian Foundation for Basic Research, project
%no.~19-01-00364.

\end{document}